\documentclass[twocolumn]{aastex63}
\usepackage{amsmath,amstext}
\usepackage[T1]{fontenc}
\usepackage{apjfonts} 
\usepackage{soul}
\usepackage{comment}
\usepackage[figure,figure*]{hypcap}

\received{~--- xxx} 
\revised{~--- xxx}
\accepted{~--- xxx}

\shorttitle{AGN in Faint Dwarf Galaxy}
\shortauthors{Paswan et al.}

\begin{document}

\title{\large An Accreting Supermassive Black Hole Buried in a Faint Dwarf Galaxy}

\correspondingauthor{Abhishek Paswan}
\email{apaswan@allduniv.ac.in}

\author{Abhishek Paswan}
\affil{Indian Institute of Astrophysics, Koramangala II Block, Bangalore 560034, India}
\affiliation{Department of Physics, University of Allahabad, Prayagraj 211002, India}

\author{Mousumi Das}
\affiliation{Indian Institute of Astrophysics, Koramangala II Block, Bangalore 560034, India}

\author{K Rubinur}
\affiliation{Institute of Theoretical Astrophysics, University of Oslo, P.O Box 1029, Blindern, 0315 Oslo, Norway}
\affiliation{National Centre for Radio Astrophysics - Tata Institute of Fundamental Research, Ganeshkhind, Pune 411007, India}

\begin{abstract}

\noindent In the last decade, there have been several discoveries of Active Galactic Nuclei (AGN) in dwarf galaxies including an AGN in an ultra-compact dwarf galaxy with a Black Hole mass $>$10$^{6}$ M$_{\odot}$. However, finding a Supermassive Black Hole (SMBH) in a dwarf Low Surface Brightness (LSB) galaxy is rare. We report the discovery of a Seyfert type-2 class AGN which is associated with a nuclear SMBH of mass $\sim$6.5 $\times$ 10$^{6}$ M$_{\odot}$ in a dwarf LSB galaxy ($\mu_{0,r}$ $>$ 23.8 mag/arcsec$^{2}$) that we denote by MJ0818+2257. The galaxy was previously thought to be an outlying emission blob around the large spiral galaxy LEDA 1678924. In our current analysis, which includes the detection of the optical counterpart of MJ0818+2257, we study its ionized gas kinematics and find that the dynamical mass within the ionized gas disk is $\sim$5.3 $\times$ 10$^{9}$ M$_{\odot}$. This is comparable to its stellar mass which is $\sim$3$\times$ 10$^{9}$ M$_{\odot}$ and suggests that MJ0818+2257 is moderately dark matter-dominated within the stellar disk. The SMBH mass to galaxy stellar mass ratio is $M_{BH}/M(*)>0.022$ which is high compared to disk galaxies. Our detection of a SMBH in a bulgeless LSB dwarf galaxy raises questions about the growth of SMBHs in low-luminosity galaxies and suggests the possibility of detecting heavy, seed black holes from early epochs in LSB dwarf galaxies in the low redshift universe.

\end{abstract}


\section{Introduction} 
\label{sec:intro}

One of the fundamental open problems of current observational cosmology is to understand the formation of the first Black Holes \citep[BHs;][]{Bellovary2011}. It is not clear how the initial seed BHs formed in the early universe, what were their properties, and what was the nature of their host galaxies. In the literature, Supermassive BH (SMBH) masses strongly correlate with many properties of the host galaxies and bulges across cosmic time, which only implies co-evolution \citep[e.g.,][]{Gebhardt2000,Ferrarese2000,Marconi2003,McConnell2013}. However, unlike massive galaxies, low luminosity dwarfs do not have bulges nor do they appear to have undergone a rich merger history. Hence, finding Active Galactic Nuclei (AGN) in these galaxies is surprising. Also, since their merger history is poor, the BHs associated with their AGN may be similar to the first seed BHs \citep{Bellovary2011}. While direct observational detection of seed BHs in massive galaxies is difficult with current capabilities, as they are at very high redshifts. The discovery of more and more very massive SMBH at very high redshifts requires either super-Eddington accretion, very large seed BHs, or both, the nearby faint dwarf galaxies are within our observational reach and can put important constraints on the connection between the host galaxies and their seed BHs \citep{Volonteri2010,Greene2012}. 

A few studies, including NGC 4395, confirm the detection of AGN having intermediate-mass BHs (IMBHs) and SMBHs in low-mass dwarf galaxies \citep[e.g.,][]{Reines2011Nature,Reines2012,Reines2016,Reines2020}. However, the detection of an AGN associated with a SMBH in a bulgeless, Low Surface Brightness (LSB) dwarf galaxy is rare and cannot be explained by conventional galaxy evolution theories or the M$_{BH}-\sigma_{*}$ relation \citep{McConnell2013}. Such a detection, however, may provide better constraints to the limiting mass of seed BHs and their growth in early epochs. This is especially important since the \textit{James Webb Space Telescope (JWST)} has detected surprisingly massive BHs in small galaxies \citep{pacucci.etal.2023}.\par

\begin{figure*}
\begin{center}
\rotatebox{0}{\includegraphics[trim=00 00 00 00,width=0.85\textwidth]{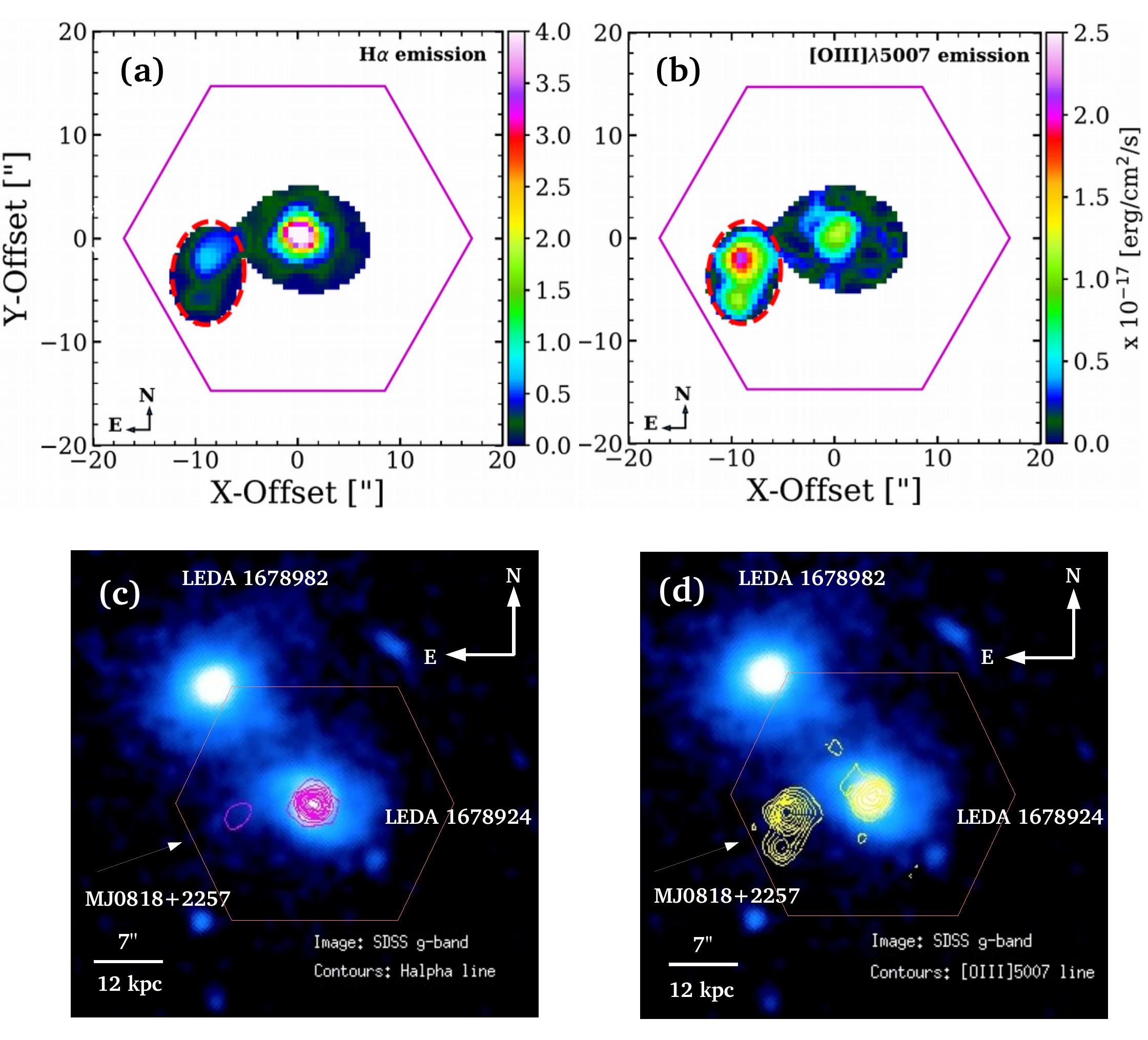}}
\caption{Panels (a) and (b) show the H$\alpha$ and [OIII]$\lambda$5007 emission line maps, respectively. In the emission line maps, the outlying emission blob (which we later name as MJ0818+2257) is marked with a red dashed ellipse. Panel (c) and (d) represent the SDSS $g$-band image overlaid with the H$\alpha$ (red) and [OIII]$\lambda$5007 (yellow) emission lines contours, indicating that there is no optical counterpart of the outlying emission blob.}
\label{fig-1}
\end{center}
\end{figure*}

Here we present the serendipitous discovery of a bulgless, LSB dwarf galaxy hosting an AGN observed with the Integral Field Unit (IFU) spectroscopy \citep{drory.etal.2015} in the Mapping Nearby Galaxies at Apache Point Observatory \citep[MaNGA;][]{Bundy2015} survey. This galaxy, which we name as MJ0818+2257,  was earlier thought to be an emission blob \citep{Omkar2019} lying outside the large spiral galaxy, LEDA 1678924, and its close lenticular companion galaxy, LEDA 1678982 (see Fig.~\ref{fig-1}a$-$d). This outlying emission blob was thought to be a Hanny’s Voorwerp (HsV) object illuminated by the AGN
activity of LEDA 1678924 \citep{Omkar2019}. However, our  investigation shows that there is an optical counterpart to the emission blob  and it is also  associated with a rotating disk of ionized gas. Hence,  the emission blob is actually a bulgeless LSB dwarf galaxy, MJ0818+2257, and the ionized gas is due to AGN activity in the galaxy. 

Throughout the paper, we have considered a flat $\Lambda$CDM cosmology with $H_{0}$ = 70 km$s^{-1}$ Mpc$^{-1}$ , $\Omega_{m}$ = 0.3, and $\Omega_{\Lambda}$ = 0.7, where $H_{0}$ represents the Hubble constant, and $\Omega_{m}$ and $\Omega_{\Lambda}$ are matter and dark energy density, respectively. All the magnitudes quoted in this paper are in the AB system \citep{Oke1974}.

\section{Data and analysis}
\label{sec:data}

\subsection{Optical imaging and integral field unit (IFU) data}

This study uses publicly available optical images from the Sloan Digital Sky Survey (SDSS) DR17, which reaches 5$\sigma$ magnitudes of 23.13, 22.7, and 20.71 in the $g, r$, and $z$ band, respectively. We also use optical images from the Dark Energy Camera Legacy Survey (DECaLS) DR9, reaching 5$\sigma$ depths of 24.0, 23.4, and 22.5 magnitudes in the $g, r$, and $z$ band, respectively. The DECaLS images have been mainly used for detecting the optical counterparts of the target source.\par

The optical IFU data used in the present work comes from the MaNGA survey. We use datacubes that have been reduced and calibrated using the Data Reduction Pipeline \citep[DRP;][]{Law2016}. In the final datacubes, the calibrated spectra have a wavelength coverage of $3600 - 10300$ \AA~ with a spectral resolution of $R$ $\sim$ 2000. These calibrated datacubes are then made into science-ready products using Data Analysis Pipeline \citep[DAP;][]{Westfall2019}. This work uses the DAP output products. The DAP uses pPXF code \citep{Cappellari2004} which fits the models to both the stellar continuum and emission line features, e.g., emission and absorption lines, identified in the spectra from each spaxel. Various gas emission line fluxes are derived from Gaussian model fits after subtracting the stellar continuum and absorption lines, and then provided in the form of 2D maps. The Galactic reddening correction to the line fluxes is applied assuming the reddening law provided by \citet{O'Donnell1994}. Throughout this work, we have used only spaxels with signal-to-noise ratio (SNR) $\geq$3.

\subsection{Radio continuum}

All the radio images are derived using archival VLA data. The 1.4-GHz radio continuum image is taken from the Faint Images of the Radio Sky at Twenty-Centimeters (FIRST) survey. This survey provides radio continuum images of sources with a typical rms of 0.15 mJy beam$^{-1}$ and an angular resolution of 5$^{\prime\prime}$ \citep{Becker1995}. Quick Look images from the Karl G. Jansky Very Large Array Sky Survey \citep[VLASS;][]{Lacy2020} are also used for our investigation. The VLASS provides S-band images centered at 3 GHz with an angular resolution of 2.5$^{\prime\prime}$. Furthermore, we have analysed higher frequency C-, X- and K-band data obtained from VLA archive (ID AG617). The observations were carried out on 14-15th September, 2001 in C-configuration centered at 4.84-GHz (C-band), 8.46-GHz (X-band) and 22.5-GHz (K-band). Each band has two channels with 50 MHz each, i.e. a total bandwidth of 100 MHz. The target region was observed for 30 sec along with the phase calibrator 0854+201 for 90-100 sec and flux calibrator 1331+305 for 30 sec. These data are reduced and analysed using the Common Astronomy Software Applications (CASA). We have followed usual data reduction methods which include flagging of bad data, setting the model for flux calibrator, initial phase calibration and then gain calibrations of calibrators. Finally, the solutions are applied to the science target. After a satisfactory calibration, the images are obtained using the {\sc{CLEAN}} task. As these are snapshot observations with a small exposure time, we are bound to perform only a single round of self-calibration to improve the images.

\subsection{Data in other bands}

Apart from the above-described data sets, we also searched for emissions associated with the target source in the far-ultraviolet (FUV) and X-ray bands using images from the Galaxy Evolution Explorer ($GALEX$) and $Chandra$ surveys, respectively. We did not detect any emission in either waveband. 

\section{Results}
\label{sec:result}

\begin{table*}
\centering
\caption{Basic parameters of the galaxies, where projected separation is wrt MJ0818+2257}.
\vspace {0.3cm}
\begin{tabular}{ccccccc} \hline
Source & RA (J2000) & Dec (J2000) & Redshift ($z$) & Projected separation & Linear size\\
       &  [hh:mm:ss] & [dd:mm:ss]  &  & [kpc] & [kpc] \\\hline

MJ0818+2257 & 08:18:50.238 & +22:57:14.86 & 0.09294 $\pm$ 0.00012 & 0 & 16.9 \\
LEDA 1678924        & 08:18:49.617 & +22:57:16.39 & 0.09192$\pm$0.00001 & 16.9 & 34.9 \\
LEDA 1678982    & 08:18:50.523 & +22:57:29.03 & 0.09239$\pm$0.00013 & 26.0 & 40.2 \\
\hline 
\label{Table1}
\end{tabular}
\end{table*}

All the images together shown in Fig.~\ref{fig-1} clearly indicate the presence of an outlying emission blob marked by the red ellipse in the emission line maps. Also, its EW(H$\alpha$) ranges over $5 - 15$~\AA. Since the ionized gas with EW(H$\alpha$) $\textless$ 3 \AA~ is defined as diffuse ionized gas (DIG) in galaxies \citep[e.g.,] []{Fernandes2010,Zhang2017,Lacerda2018}, therefore, this outlying emission blob rules out the possibility that it is contaminated by DIG.\par

\subsection{Spatially-resolved 2D-BPT diagram}

\begin{figure*}[h!!!!]
\rotatebox{0}{\includegraphics[trim=10 00 00 00,width=1.0\textwidth]{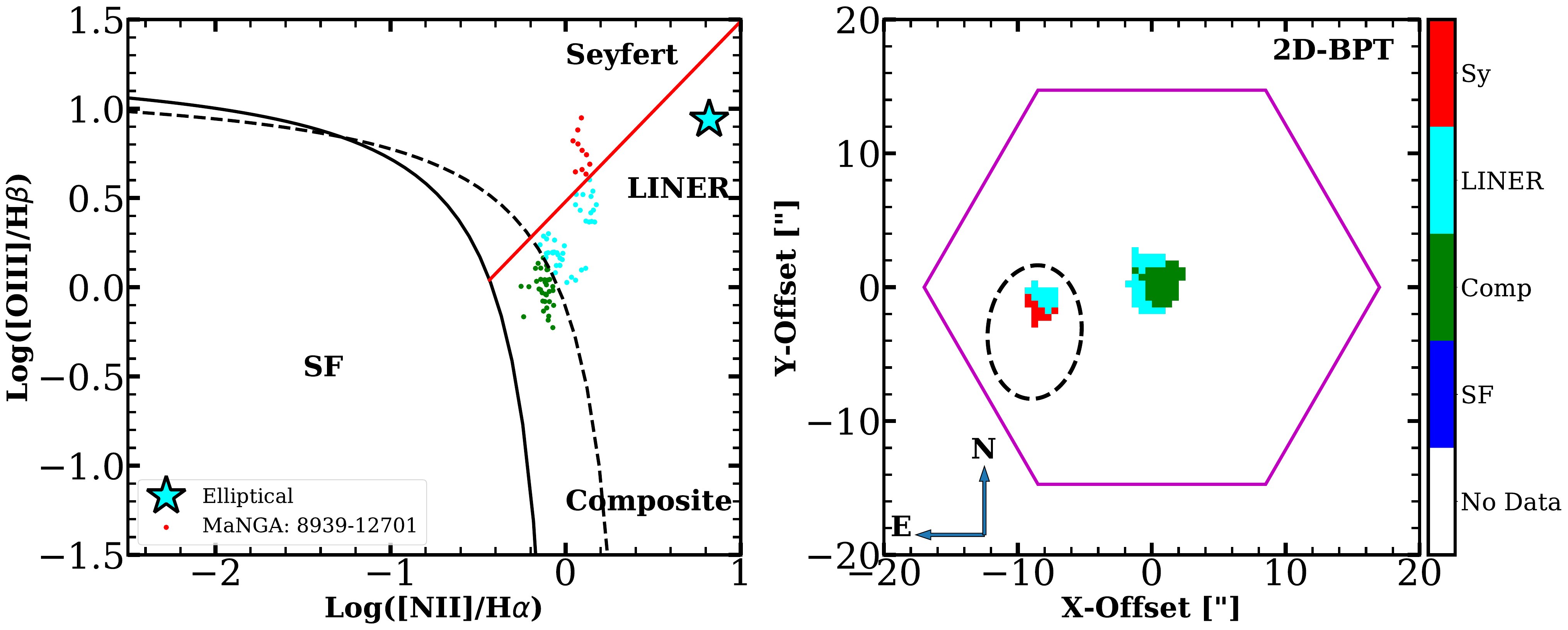}}
\caption{Left: The BPT diagram labeled according to the different regions related to star formation (SF), AGN (Seyfert or LINER) and composite (AGN+SF) activities. Spaxels from LEDA 1678924 and MJ0818+2257 are denoted by dots, while the star symbol represents the data point for LEDA 1678982 (see Fig.~\ref{fig-1}c and d). Right: the color-coded spatially-resolved 2D-BPT diagram corresponds to the left figure.}
\label{fig-2}
\end{figure*}

We use the Baldwin, Phillips, \& Terlevich \citep[BPT;][]{Baldwin1981} diagnostic diagram to explore the nature of the outlying emission blob, including LEDA 1678924 and LEDA 1678982. The MaNGA IFU data provides spatially resolved emission line data for sources in the MaNGA hexagonal footprint as shown in the right panel of Fig.~\ref{fig-2}. In the case of LEDA 1678982, exploring its spatially-resolved 2D-BPT diagnostic is impossible due to the unavailability of IFU data. From our derived spatially-resolved 2D-BPT diagnostic, we find that composite and LINER-like AGN activities are the primary sources of the ionized gas in the central region of LEDA 1678924. However, in the case of the outlying emission blob, it is due to pure AGN activity composed of  LINER and Seyfert-like activity. The nucleus of LEDA 1678982 indicates LINER-like ionization as shown by the star symbol in the left panel of Fig.~\ref{fig-2}, which is derived using 3" fiber slit spectrum observed in the SDSS survey. 

\subsection{Detection of the optical counterpart of outlying emission blob}

\begin{figure*}
\begin{center}
\rotatebox{0}{\includegraphics[trim=00 00 00 00,width=1.0\textwidth]{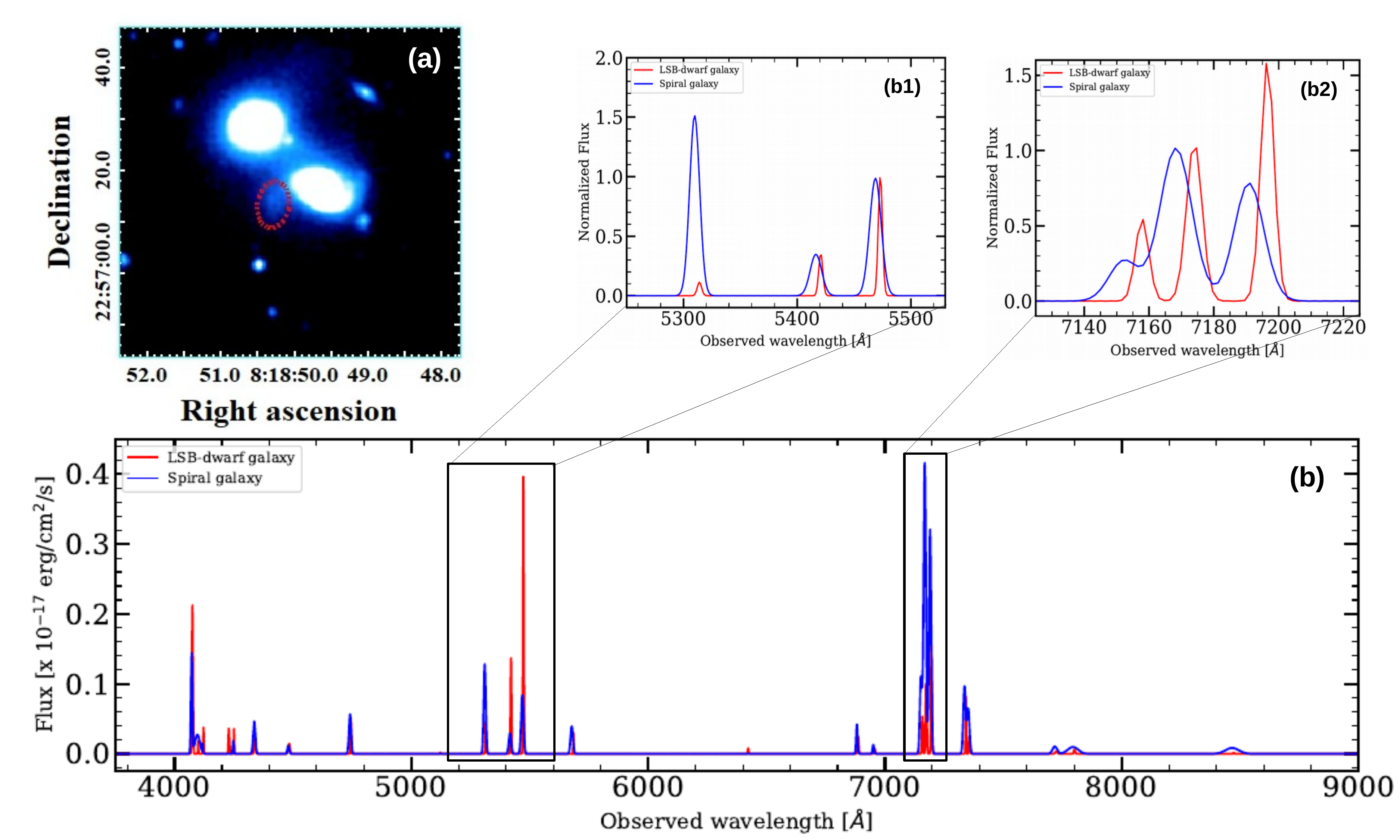}}
\caption{Panel (a) shows the DECaLS $g-$band image of MaNGA target 1-383997 i.e., LEDA 1678924. The detection of optical counterpart of MJ0818+2257 is marked by red ellipse. Panel (b) represents the stellar continuum and absorption subtracted observed spectra of LEDA 1678924 (i.e., spiral galaxy) and outlying emission blob (i.e., LSB dwarf galaxy). Panel (b1) shows the zoomed view of H$\beta$ [OIII]$\lambda$4959,5007 emission lines. Similarly, panel (b2) presents the zoomed view of [NII]$\lambda$6548,6583 and H$\alpha$ emission lines.}
\label{fig-3}
\end{center}
\end{figure*}

In Fig.~\ref{fig-3}(a), the DECaLS $g$-band image available via the recent data release (DR9) clearly shows the detection of an optical counterpart of the outlying emission blob. It is unlikely that this optical detection is due to strong gas line emission falling in the filter bands because the strongest emission line [OIII]$\lambda$5007 and other lines around this wavelength fall outside the $g$-band (see Fig.~\ref{fig-7} in the Appendix). Moreover, a very faint detection is also noticed in the $r$, $i$ and $z$-band images taken from the DECaLS survey (see Fig.~\ref{fig-6} in the Appendix).\par 
Fig.~\ref{fig-3}(b) shows the continuum and stellar absorption subtracted emission line spectra taken at the center of the outlying emission blob (red) and LEDA 1678924 (spiral galaxy; blue), whose sub-images with normalized flux around the rest-wavelengths of [OIII]$\lambda$4958,5007, H$\beta$ and H$\alpha$, [NII]$\lambda$6548,6583 emission lines are shown in Fig.~\ref{fig-3}(b1) and (b2) respectively. These emission lines indicate that the outlying emission blob and LEDA 1678924 are at different redshifts of 0.09192$\pm$0.00001 and 0.09294$\pm$0.00012, respectively (see Table.~\ref{Table1}). The third system in the field, LEDA 1678982, is at a  redshift of 0.09239$\pm$0.00013, derived
using 3" fiber slit spectrum observed in the SDSS survey. This implies that all the three systems, within the errors, are at different redshifts, and the outlying emission blob is the most redshifted relative to LEDA 1678924 and LEDA 1678982. In Table~\ref{Table1}, it can be seen that LEDA 1678924 and LEDA 1678982 are separated by projected distances of $\sim$ 16.9 and $\sim$ 26 kpc respectively from the outlying emission blob. These results reveal the possibility of the existence of a very faint, low-mass, dwarf galaxy candidate (hereafter named as MJ0818+2257) contributing to the stellar emission at the location of the outlying emission blob. 

\subsection{Kinematics of emission blob}

\begin{figure*}
\begin{center}
\rotatebox{0}{\includegraphics[trim=00 25 35 00,width=0.85\textwidth]{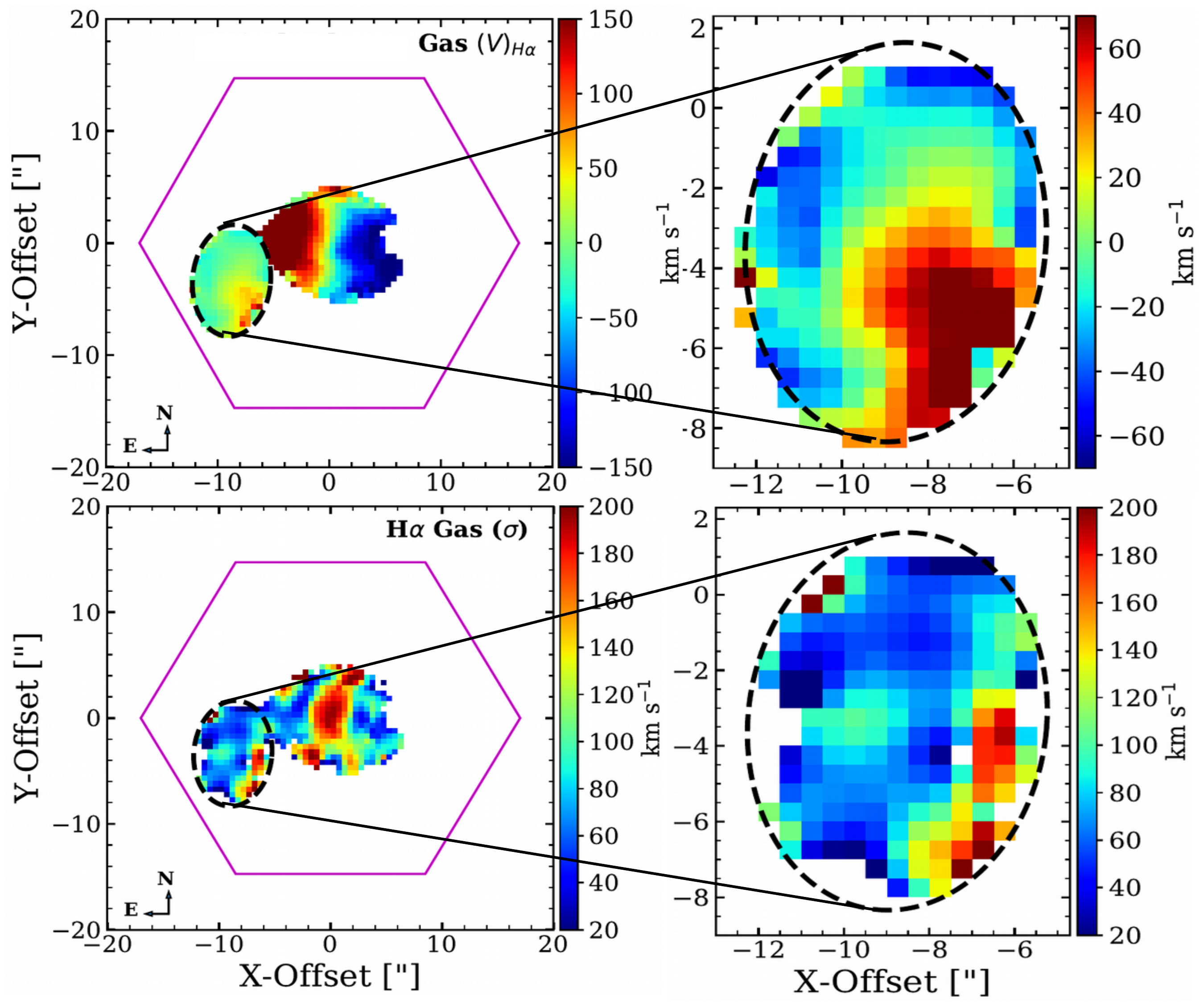}}
\caption{The rotation velocity of ionized gas derived using H$\alpha$ emission line (top left panel), and top right panel represents its zoomed in view  centered on the central velocity of MJ0818+2257. Similarly, bottom left panel shows the velocity dispersion, and bottom right panel represents its zoomed in view. In each panel, the location of MJ0818+2257 is marked by black dashed ellipse.}
\label{fig-4}
\end{center}
\end{figure*}

The gas kinematics based on the H$\alpha$ emission line of LEDA 1678924 shows that LEDA 1678924 has a regularly rotating disk with a velocity field typical of disk galaxies. However, the velocity field shows a discontinuity at the location of MJ0818+2257, which is unusual if we assume both galaxies are at the same redshift (for more detail see Appendix). In fact, the observed velocities clearly show a dual distribution. This suggests that MJ0818+2257 and LEDA 1678924 are kinematically decoupled, implying that they are two independent systems. As observed in the previous section, MJ0818+2257 and LEDA 1678924 are at different redshifts. Hence we use their corresponding systematic velocities to derive their individual velocity fields as presented in Fig.~\ref{fig-4}. Here the derived kinematics is corrected for the inclination angle ($i$) of galaxy, following the relation $V_{rot}=V^{obs}_{los}/Sin(i)$. The inclination angle for MJ0818+2257 is estimated using $i=Cos^{-1}$(b/a), where a and b represent the sizes of major and minor axes of outer ellipse, respectively. Since the stellar emission lines from MJ0818+2257 are too faint to detect using MaNGA IFU data, it is not possible to derive the stellar kinematics.\par

In Fig.~\ref{fig-4}, it can be noticed that MJ0818+2257 shows a clear gas rotation, similar to the rotating dwarf galaxies studied in \citet[][]{Anil2014}. Also, the gas velocity dispersion ranges over $50-100$ km~s$^{-1}$, which is again similar to typical rotating dwarf galaxies \citep[e.g.,][]{Anil2014,Paswan2022}. This finding safely rules out the possibility that MJ0818+2257 is an outlying emission blob originated due to the HsV process \citep[see][]{Jozsa2009} or outflow ejections via AGN sources, which was previously reported by \citet{Omkar2019}. Overall, our results show that MJ0818+2257 is an independent, rotating dwarf galaxy. Using the maximum gas rotation velocity ($v$)\textbf{,} which is $\sim$ 52 km~s$^{-1}$, we estimate the dynamical mass ($M_{dyn}\sim v^{2}R/G$) of MJ0818+2257 to be $\sim$ 5.3 $\times$ 10$^{9}$ M$_{\odot}$, within its disk radius ($R$) of $\sim$ 8.5 kpc.\par

\subsection{Views from multi-frequency radio continuum emissions}

\begin{figure*}
\rotatebox{0}{\includegraphics[trim=80 00 00 00,width=1.0\textwidth]{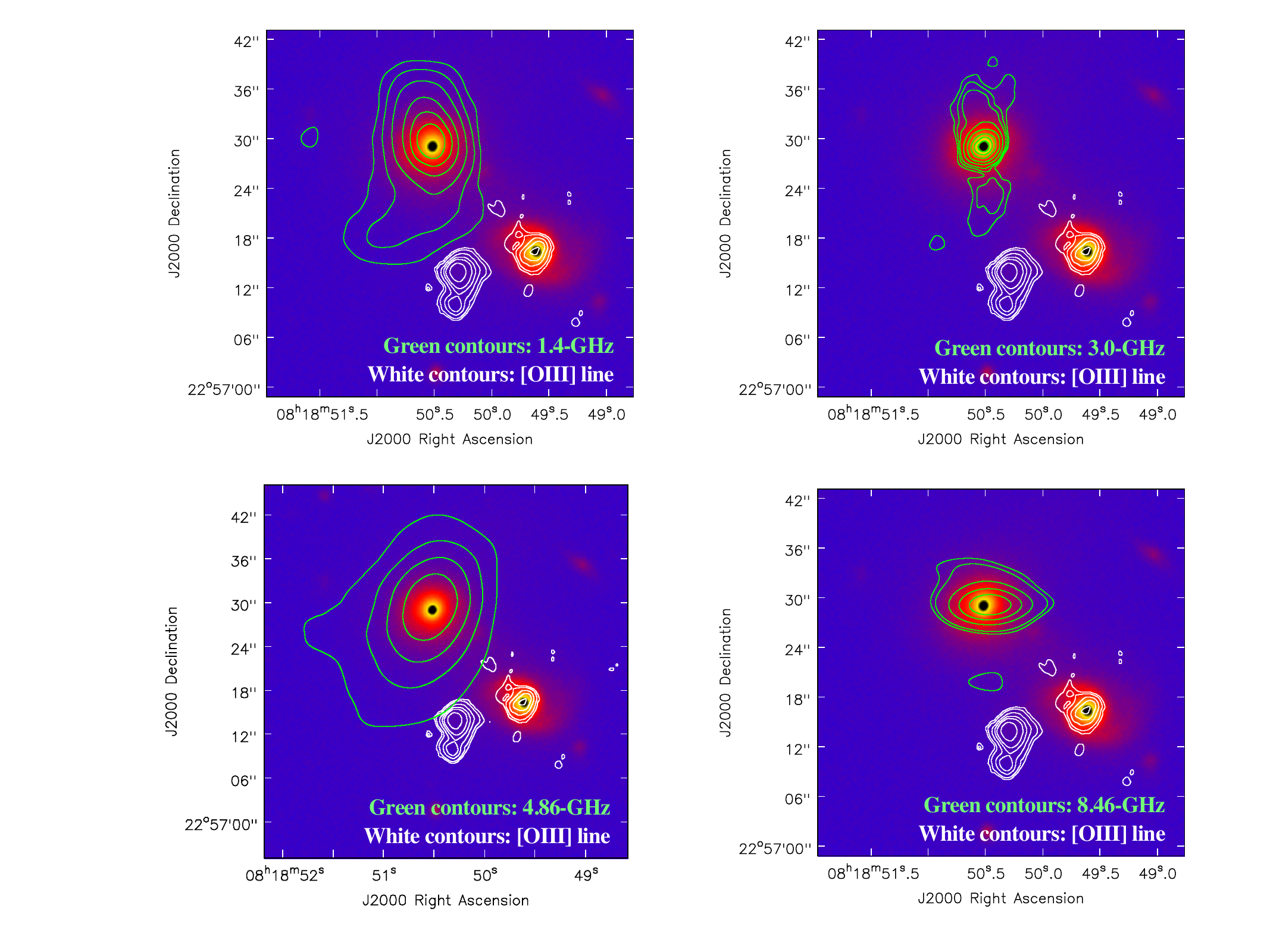}}
\caption{The radio continuum contours as shown by green solid lines at different frequencies that include 1.4-GHz (top left), 3-GHz (top right), 4.86-GHz (bottom left) and 8.46-GHz (bottom right) overlaid on optical g-band image from the DECaLS survey. In each panel, the white contours represent the MaNGA [OIII]$\lambda$5007 emission line at the location of MJ0818+2257 and the spiral galaxy. Here north and east are up and left, respectively.}
\label{fig-5}
\end{figure*}

In Fig.~\ref{fig-5}, we show the optical $r$-band image overlaid with radio continuum contours (green) at 1.4, 3, 5, and 8.5-GHz. In each panel of Fig.~\ref{fig-5}, white contours represent the [O{\sc iii}]$\lambda$5007 emission line from LEDA 1678924 and MJ0818+2257 observed in the MaNGA survey. Note that the [O{\sc iii}]$\lambda$5007 emission line is also observed from LEDA 1678982, but it is not possible here to show its extent in the form of contours due to slit (3'' fiber) observations in the SDSS survey. Interestingly, it is noticeable that radio continuum emissions at different radio frequencies are seen from LEDA 1678982 only, while the three systems (i.e., MJ0818+2257, LEDA 1678924, and LEDA 1678982) are optically identified as hosting AGN sources (see BPT diagram). In fact, a radio jet-like structure from LEDA 1678982 can be observed in the 3-GHz radio band. However, there is a lack of close association between radio emission from LEDA 1678982 and MJ0818+2257, unlike the previous case of IC 2497, whose radio continuum emission was found closely associated with an HsV object \citep{Jozsa2009}. Such a close association is essential for the origin of HsV objects because a plasma jet associated with AGN helps to clear a path through ISM/IGM towards the nebulosity. Also, such a scenario is not even seen between LEDA 1678924 and MJ0818+2257. Thus, our result supports the fact that MJ0818+2257 is an independent dwarf galaxy.\par 

\section{Observed nature of MJ0818+2257}

Our analyses reveal the following, (i) the detection of the optical counterpart of MJ0818+2257 which reveals a well-shaped dwarf galaxy-like morphology \citep[see Fig.~\ref{fig-3}, and is unlike the HsV object around IC 2497;][]{Schawinski2010}; (ii) a rotating gaseous disk associated with MJ0818+2257 \citep[see Fig.~\ref{fig-4} which is similar to M60-UCD1;][]{Anil2014}. Both facts strongly indicate that MJ0818+2257 is an independent low-luminosity dwarf galaxy that hosts a Seyfert-type AGN source (see BPT diagram in Fig.~\ref{fig-2}). The galaxy has a diameter of $\sim$ 16.9 kpc. Its central surface brightness ($\mu_{0,r}$), estimated using the DeCALS $r$-band image, is found to be fainter than 23.8 mag/arsec$^{2}$ (see Appendix). The central surface brightness-based criteria used for classifying a galaxy into a Low Surface Brightness (LSB) galaxy, given by \citet{Adami2006}, indicates that MJ0818+2257 is an LSB dwarf galaxy. Such galaxies have low stellar surface densities and are metal-poor but very gas-rich \citep[e.g.,][]{honey.etal.2018}. The blue color ($g - r$ $\textless$ 0.8) of MJ0818+2257 indeed confirms that it is metal-poor, and the strong H$\alpha$ and [O {\sc iii}]$\lambda$5007 emissions suggest that it is rich in ionized gas. The stellar mass, derived from the DeCALS $r$ and $z-$band images is $\textless$ 3 $\times$ 10$^{9}$ M$_{\odot}$ (see Appendix).\par
Since MJ0818+2257 hosts an AGN, we derived its BH mass using the [O{\sc iii}] $\lambda$5007 bolometric nuclear luminosity of MJ0818+2257, assuming the Eddington ratio of a typical Seyfert type-2 class AGN (see see Appendix). The mass of the BH hosted in MJ0818+2257 is found to be $\sim$6.5 $\times$ 10$^{6}$ M$_{\odot}$ which falls in the SMBH category. 

\section{Discussion and conclusion}

In the literature, SMBHs are commonly found in massive galaxies that have a significant bulge component \citep{McConnell2013}. But recent discoveries show that massive BHs can be found in low-luminosity dwarf galaxies \citep{Reines2013,Baldassare2015} and bulgeless disk galaxies as well \citep{Filippenko2003}. For example, using optical SDSS spectroscopy, radio and X-ray data, studies by Reines et al. \citep[e.g.,][]{Reines2012,Reines2013,Reines2016,Reines2020} and others \citep{Greene2007,Greene2012} reported more than a hundred dwarf galaxies that host BHs in the mass range of 10$^{5}$ $\leq$ M$_{BH}$ $\leq$ 10$^{6}$ M$_{\odot}$. Recently, using MaNGA IFU data, \citet{Mezcua2020} also reported a total of thirty seven AGN dwarf galaxies hosting IMBHs (M$_{BH}$ $\sim$ 10$^{5}$ M$_{\odot}$), which were previously missed out with slit spectroscopy due to their off nuclear AGN activities. Also, a SMBH ($>$ 10$^6$~M$_\odot$) was detected in the ultra-compact dwarf galaxy M60-UCD1 using NIR-IFU observations \citep{Anil2014}. \par

Considering these past studies, our discovery of an AGN and a SMBH ($\sim$10$^6$ M$_\odot$) in the dwarf LSB galaxy  MJ0818+2257, is important for several reasons. First, it shows the importance using ionized emission lines for the detection of accreting massive BHs in low luminosity sources such as MJ0818+2257. It also shows that IFU facilities such as MaNGA play a critical role for discovering rare low luminosity galaxies that can be detected only from emission line observations, and not only optical imaging observations. For example, [OIII] or H$\alpha$ emission in combination with UV continuum imaging have been used to detect faint galaxies even in our nearby universe \citep{yadav.etal.2022}. Secondly, SMBHs in dwarfs are important for establishing the scaling relations between the mass of BHs and their host galaxies at the lowest mass scales \citep[e.g.,][]{Kormendy2013,Bentz2018,Schutte2019}. The third reason is that it raises the possibility of detecting large populations of SMBHs in dwarf LSB galaxies at high redshifts using IFU facilities such as the $JWST$. This is especially important because large populations of such dwarf galaxies are expected to exist at early epochs and may be important for re-ionization of the universe \citep{atek.etal.2024}. MJ0818+2257 is thus a good example of the low redshift counterpart of such dwarf galaxies. 

Another important fact is that the SMBH in MJ0818+2257 may represent a seed BH from early epochs in the universe. This is because the host galaxy is an LSB dwarf and such galaxies do not show signatures of a rich merger history that would have led to the growth of their nuclear BH. LSB galaxies are known to have low stellar surface densities, low metallicities and low star formation rates. For example, giant LSB galaxies that also have diffuse stellar disks and have not undergone many mergers, have relatively low mass SMBHs (10$^{5}$ - 10$^{7}$ M$_{\odot}$) that lie below the $M-\sigma$ relation for low-z galaxies \citep{subramanian.etal.2016}. There are also cosmological models and simulations of BH evolution that suggest that low-mass dwarf galaxies are potential systems hosting pristine BHs that are the seeds for SMBHs in more massive galaxies \citep{Angles-Alcazar2017,Habouzit2017}. The SMBH mass to stellar mass ratio in MJ0818+2257 i.e. M(BH)/M(*)$>$0.022 is slightly higher than the local value of $\sim$0.001 (see Appendix A) \citep{dayal.2024}.  Thus the SMBH in MJ0818+2257 may represent a massive seed BHs from early epochs that has remained relatively pristine due to a lack of mergers in its evolutionary history.

There are mainly three processes that lead to the formation of massive seed BHs at early epochs. They include the direct cloud collapse models that form heavy seed BHs, the evolution of $\sim$100M${_\odot}$ population III stars into SMBHs via galaxy mergers and the evolution of dense stellar clusters at early epochs \citep{cammelli.etal.2024}. But since MJ0818+2257 is a bulgeless, metal-poor dwarf, the origin of its SMBH can best be explained only via the direct collapse of a pristine gas cloud, which is similar to the recently discovered X-ray quasar, UHZ1, at $z~\sim$ 10, that has been confirmed to harbor a heavy seed BH using $JWST$/NIRCam observations \citep{Bogdan2023}. Thus, dwarfs such as MJ0818+2257 are ideal low-z laboratories for placing constraints on the nature and mass of seed BHs, as well as provide inputs for modelling their growth.


\section*{Acknowledgment}

MD acknowledges the support of the Science and Engineering Research Board (SERB)
Core Research Grant CRG/2022/004531 for this research. We thank the anonymous referee for constructive suggestions that have greatly improved the quality of the
paper. 

This research has made use of the SAO/NASA Astrophysics Data System (ADS) operated by the Smithsonian Astrophysical Observatory (SAO) under a NASA grant. Funding for the Sloan Digital Sky Survey IV has been provided by the Alfred P. Sloan Foundation, the U.S. Department of Energy Office of Science, and the Participating Institutions. SDSS-IV acknowledges support and resources from the Center for High Performance Computing at the University of Utah. \par

SDSS-IV is managed by the Astrophysical Research Consortium for the Participating Institutions of the SDSS Collaboration, including the Brazilian Participation Group, the Carnegie Institution for Science, Carnegie Mellon University, Center for Astrophysics Harvard \& Smithsonian, the Chilean Participation Group, the French Participation Group, Instituto de Astrofísica de Canarias, The Johns Hopkins University, Kavli Institute for the Physics and Mathematics of the Universe (IPMU)/University of Tokyo, the Korean Participation Group, Lawrence Berkeley National Laboratory, Leibniz Institut für Astrophysik Potsdam (AIP), Max-Planck-Institut für Astro- nomie (MPIA Heidelberg), Max-Planck-Institut für Astrophysik (MPA Garching), Max-Planck-Institut für Extraterrestrische Physik (MPE), National Astronomical Observatories of China, New Mexico State University, New York University, University of Notre Dame, Observatário Nacional/MCTI, The Ohio State University, Pennsylvania State University, Shanghai Astronomical Observatory, United Kingdom Participation Group, Universidad Nacional Autónoma de México, University of Arizona, University of Colorado Boulder, University of Oxford, University of Portsmouth, University of Utah, University of Virginia, University of Washington, University of Wisconsin, Vanderbilt University, and Yale University.\par

The Legacy Surveys consist of the Dark Energy Camera Legacy Survey (DECaLS), the Beijing-Arizona Sky Survey and the Mayall z-band Legacy Survey. DECaLS, BASS. The data is obtained at the Blanco telescope, Cerro Tololo Inter-American Observatory, NSF’s NOIRLab; the Bok telescope, Steward Observatory, University of Arizona; and the Mayall telescope, Kitt Peak National Observatory, NOIRLab. Pipeline processing were supported by NOIRLab and the Lawrence Berkeley National Laboratory (LBNL). The Legacy Surveys project is honored to be permitted to conduct astronomical research on Iolkam Du’ag (Kitt Peak), a mountain with particular significance to the Tohono O’odham Nation.\par

This project used data obtained with the Dark Energy Camera (DECam), which was constructed by the Dark Energy Survey (DES) collaboration. Funding for the DES Projects has been provided by the U.S. Department of Energy, the U.S. National Science Foundation, the Ministry of Science and Education of Spain, the Science and Technology Facilities Council of the United Kingdom, the Higher Education Funding Council for England, the National Center for Supercomputing Applications at the University of Illinois at Urbana-Champaign, the Kavli Institute of Cosmological Physics at the University of Chicago, Center for Cosmology and Astro-Particle Physics at the Ohio State University, the Mitchell Institute for Fundamental Physics and Astronomy at Texas A\&M University, Financiadora de Estudos e Projetos, Fundacao Carlos Chagas Filho de Amparo, Financiadora de Estudos e Projetos, Fundacao Carlos Chagas Filho de Amparo a Pesquisa do Estado do Rio de Janeiro, Conselho Nacional de Desenvolvimento Cientifico e Tecnologico and the Ministerio da Ciencia, Tecnologia e Inovacao, the Deutsche Forschungsgemeinschaft and the Collaborating Institutions in the Dark Energy Survey. The Collaborating Institutions are Argonne National Laboratory, the University of California at Santa Cruz, the University of Cambridge, Centro de Investigaciones Energeticas, Medioambientales y Tecnologicas-Madrid, the University of Chicago, University College London, the DES-Brazil Consortium, the University of Edinburgh, the Eidgenossische Technische Hochschule (ETH) Zurich, Fermi National Accelerator Laboratory, the University of Illinois at Urbana-Champaign, the Institut de Ciencies de l’Espai (IEEC/CSIC), the Institut de Fisica d’Altes Energies, Lawrence Berkeley National Laboratory, the Ludwig Maximilians Universitat Munchen and the associated Excellence Cluster Universe, the University of Michigan, NSF’s NOIRLab, the University of Nottingham, the Ohio State University, the University of Pennsylvania, the University of Portsmouth, SLAC National Accelerator Laboratory, Stanford University, the University of Sussex, and Texas A\&M University.\par

BASS is a key project of the Telescope Access Program (TAP), which has been funded by the National Astronomical Observatories of China, the Chinese Academy of Sciences, and the Special Fund for Astronomy from the Ministry of Finance. The BASS is also supported by the External Cooperation Program of Chinese Academy of Sciences, and Chinese National Natural Science Foundation.\par

The Legacy Survey team makes use of data products from the Near-Earth Object Wide-field Infrared Survey Explorer (NEOWISE), which is a project of the Jet Propulsion Laboratory/California Institute of Technology. NEOWISE is funded by the National Aeronautics and Space Administration.\par

The Legacy Surveys imaging of the DESI footprint is supported by the Director, Office of Science, Office of High Energy Physics of the U.S. Department of Energy under Contract No. DE-AC02-05CH1123, by the National Energy Research Scientific Computing Center, a DOE Office of Science User Facility under the same contract; and by the U.S. National Science Foundation, Division of Astronomical Sciences under Contract No. AST-0950945 to NOAO.\par

The National Radio Astronomy Observatory is a facility of the National Science Foundation operated under cooperative agreement by Associated Universities, Inc.

\appendix

\section{Optical photometry and stellar mass estimation}
\begin{figure*}
\begin{center}
\rotatebox{0}{\includegraphics[trim=00 00 00 00,width=1.0\textwidth]{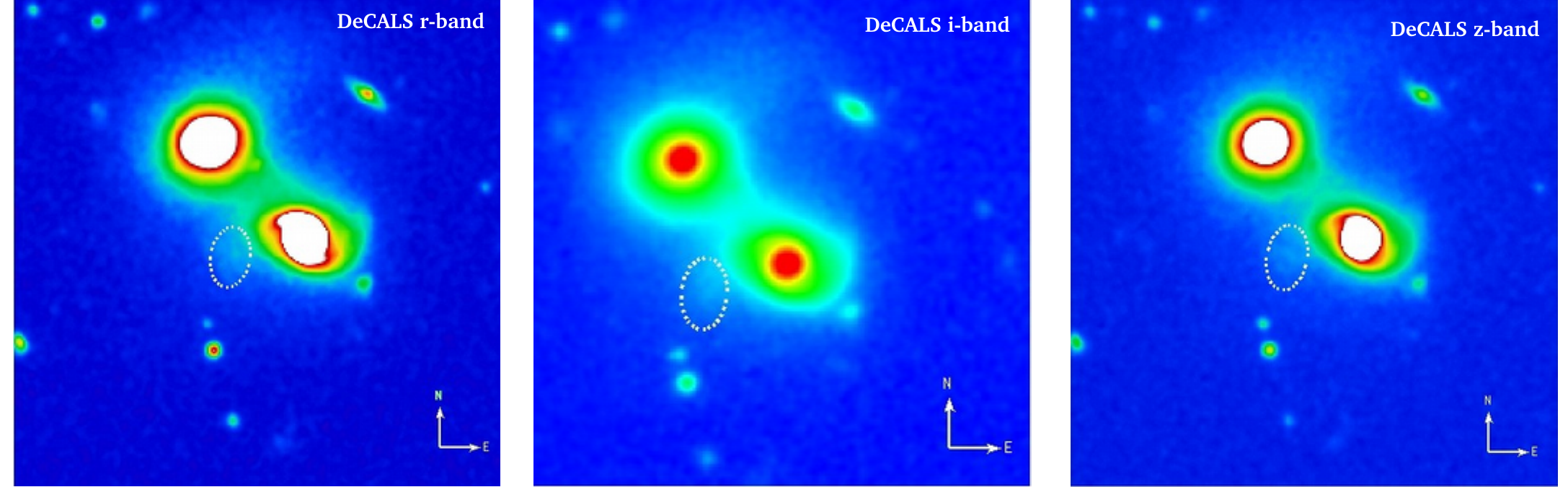}}
\caption{Deep optical $r$ (left), $i$ (middle) and $z-$band (right) images of LEDA 1678924 and LEDA 1678982, taken from the DeCALS survey. In each panel, a dotted ellipse represents the location of MJ0818+2257.}
\label{fig-6}
\end{center}
\end{figure*}

\begin{figure*}
\begin{center}
\rotatebox{0}{\includegraphics[trim=00 00 00 00,width=0.65\textwidth]{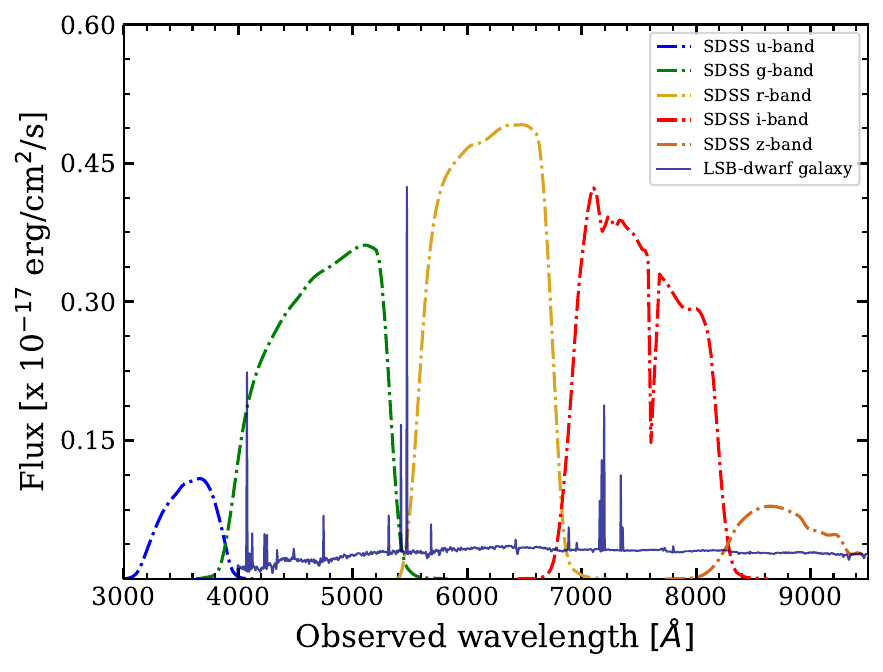}
}
\caption{The observed spectrum at the center of the outlying emission blob (i.e., MJ0818+2257) overlaid with the five SDSS filter bands, showing contaminations of emission line features with the filter bands.}
\label{fig-7}
\end{center}
\end{figure*}

The optical counterpart of MJ0818+2257 is best detected in the DeCALS $g-$band image (see Fig.~\ref{fig-3}), but is also faintly detected in the $r,~i$ and $z-$band images (see Fig.~\ref{fig-6}). Furthermore, Fig.~\ref{fig-7} justifies that the emission in most of these bands is mainly dominated by stellar continuum emission and is not due to strong nebular emission lines. However, some faint light contamination due to LEDA 1678924 cannot be ignored in these broadband filters.\par 

The optical detection of MJ0818+2257 over multiple bands offers an opportunity to constrain its stellar mass. There are several methods to estimate the stellar mass of a galaxy, and most of these methods involve the stellar population's mass-to-light ratio (M/L). Although the M/L obtained by fitting the stellar population synthesis (SPS) models to the spectral energy distribution \citep[SED;][]{Walcher2011,Conroy2013,Courteau2014} of a galaxy is the most reliable one, in the absence of a broad span of an SED, the color-based M/L estimates are widely used to establish several calibrations that provide the stellar mass of a galaxy \citep[e.g.,][]{Bell2001,Bell2003,Gallazzi2009,McGaugh2014}. These calibrations are made after taking into account the underlying stellar population, dust attenuation, and chemical evolution. Note that color-based stellar mass (or M/L) does not deviate much from that of the SED-based estimates \citep{Roediger2015}. Using a mock galaxy sample, \citet{Roediger2015} have demonstrated that the stellar masses derived based on optical color and SED fitting are consistent within a scatter of 0.2 dex and yield similar results when these methods are applied to galaxies having their stellar masses in the range of $\sim$ 10$^{8}$ $-$ 10$^{11}$ M$_{\odot}$. \par

Since MJ0818+2257 is detected in the DeCALS $g,~r,~i$ and $z-$band only, this is not enough to construct the SED and we use the color-based approach to estimate the M/L by following the calibration given below \citep{Bell2003}.  

\begin{equation}
    Log(M/L) = a_{\lambda} + (b_{\lambda} \times color),
\end{equation}

\noindent where M/L is in solar units, and a$_{\lambda}$ and b$_{\lambda}$ are color-based coefficients \citep{Bell2003}. The stellar mass of MJ0818+2257 is estimated using the $z-$band luminosity and ($r-z$) color by following the above calibration. For this combination of luminosity and color, the values of a$_{\lambda}$ and b$_{\lambda}$ are $-~ 0.041$ and 0.463 \citep{Bell2003}. It is to be noted that the above calibration yields a wavelength-dependent error in the range of $\sim$ $0.1 - 0.3$ dex for the optical bands. However, this error is observed to be minimal when we use the redder band luminosity because it is less sensitive to extinction and the age of the stellar population compared to bluer bands \citep{Tremonti2004}. We therefore choose $z-$band luminosity, which is also less contaminated by strong nebular emission lines. The $z-$band luminosity is derived using the relation, $L = 10^{0.4 \times (4.50-M_z)}~L_{\odot}$, where M$_{z}$ represents the absolute magnitude of the source in $z-$band, and is estimated using the relation $M_{z}=m_{z}-5Log(D_{L}-1)-A_{\lambda}$, where m$_{z}$ and D$_{L}$ are the apparent magnitude in $z-$band and the luminosity distance of the source, respectively and A$_{\lambda}$ represents the dust extinction correction. In the latter relation, we take the value of A$_{\lambda}$ as zero, because the observed flux ratio f$_{H\alpha}$/f$_{H\beta}$ (i.e., Balmer decrement) at the location of MJ0818+2257, derived using the MaNGA IFU data, is found to be less than the expected theoretical value of 2.86. Such low values are often associated with extremely low reddening, as has also been noticed in several past studies \citep{Ramya2009,Gunawardhana2013,Paswan2018}.\par

In order to calculate the stellar mass of MJ0818+2257, we first estimated its apparent magnitude (m$_{z}$) derived via photometry within the aperture (3.1" $\times$ 4.9") as shown in Fig.~\ref{fig-3}(a). The value of m$_{z}$ is found as $\sim$ 18.80. Similarly, the apparent magnitudes of MJ0818+2257 in the DeCALS $g$ and $r-$band images is estimated to be $\sim$ 20.24 and $\sim$ 19.44, respectively. Consequently, the values of ($g-r$) and ($r-z$) are found to be $\sim$ 0.8 and $\sim$ 0.64, respectively. Using these estimates of magnitudes and colors and following the relations described earlier, we find that the stellar mass of MJ0818+2257 as $\sim$ 3 $\times$ 10$^{9}$ M$_{\odot}$. While the magnitudes of MJ0818+2257 are derived within the aperture 3.1" $\times$ 4.9", we derive its central surface brightness ($\mu_{0,r}$ $\sim$ 23.8 mag/arcsec$^{2}$) using the aperture of 1" located at the center. This aperture is particularly selected as 1" because it is comparable to the PSF of DeCALS images. It is important to note here that the derived values of magnitudes, colors, stellar mass, and central surface brightness must be taken as their upper limits because diffuse light contamination within the selected apertures due to  LEDA 1678924 (i.e., a galaxy in the projected vicinity of MJ0818+2257) is not removed. 

\section{Observed decoupled gas kinematics}

\begin{figure}
\begin{center}
\rotatebox{0}{\includegraphics[trim=00 00 10 00,width=0.5\textwidth]{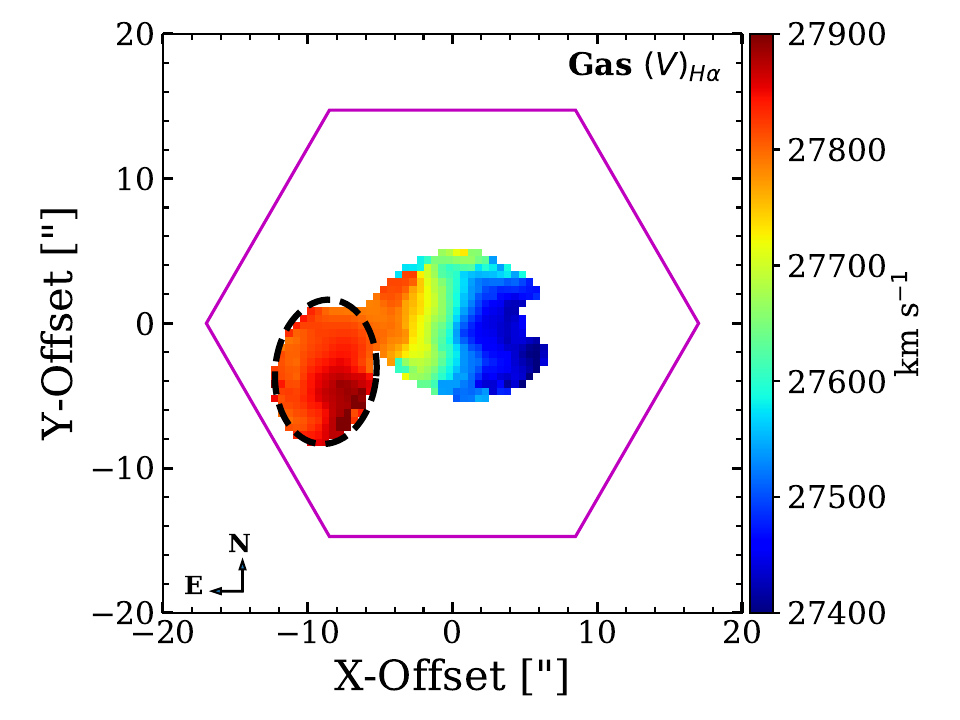}}
\rotatebox{0}{\includegraphics[trim=00 00 50 00,width=0.45\textwidth]{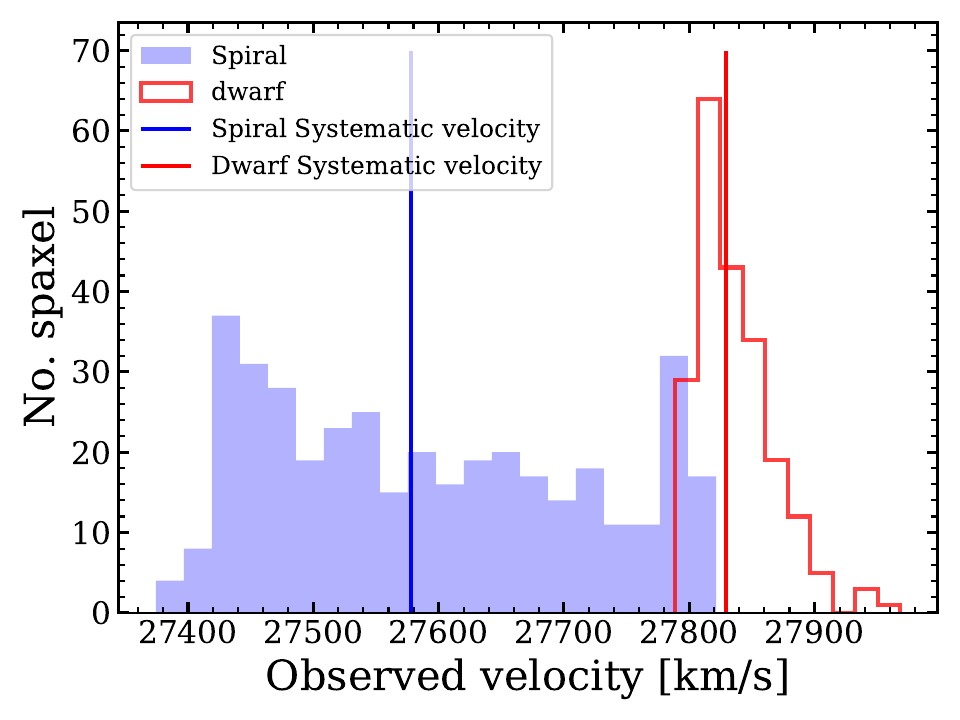}}
\caption{Left: the observed velocity map for LEDA 1678924 (i.e., spiral galaxy) and LSB dwarf galaxy (i.e., outlying emission blob). Right: their corresponding histogram velocity distributions, and their respective systematic velocities are shown by vertical lines.}
\label{fig-8}
\end{center}
\end{figure}

Using H$\alpha$ emission line, in the left panel of Fig~\ref{fig-8}, we present the observed line-of-sight velocity ($V_{los}$) map for both LEDA 1678924 and outlying emission blob (marked with dashed ellipse). In this figure, it can be noticed that the observed $V_{los}$ corresponding to the extent of LEDA 1678924 ranges over $\sim$ $27400 - 27800$ km~s$^
{-1}$ having a smooth gradient from its outer west side to east side, with a median value of $\sim$ 27580 km~s$^{-1}$ at its central region. And, it is $\sim$ $27790 - 27920$ km~s$^{-1}$ over the outlying emission blob, with a median  value of $\sim$ 27855 km~s$^{-1}$ at its central region. The velocity histogram is also shown in the right panel of Fig~\ref{fig-8}. The observed $V_{los}$ of outlying emission blob indicates a discontinuity against a regular $V_{los}$ of LEDA 1678924 (i.e., a spiral rotating disk galaxy). In the right panel of Fig~\ref{fig-8}, one can see a clear two different $V_{los}$ distribution whose central velocities significantly differ to each other. Such a discontinuity in the velocities of two objects can be only seen if they are kinematically decoupled or external to each other. This also implies that both the objects have their own different systematic velocities as shown by the two vertical lines in the right panel of Fig~\ref{fig-8}. Thus, LEDA 1678924 and outlying emission blob can be treated as two separate systems.   

\section{Black hole mass estimation}

There are several methods to calculate the BH mass in a galaxy \citep[e.g.,][]{Marconi2003,Kormendy2013,McConnell2013}, among which a method that uses the width of the broad component of the H$\alpha$ emission line is widely used in the literature \citep[e.g.,][]{Greene2007,Ramya2011}. Since we do not observe the H$\alpha$ broad component in the case of our bulgeless LSB dwarf galaxy, MJ0818+2257, we therefore choose an alternate method that estimates the BH mass. We derive the BH mass of MJ0818+2257 using its Eddington luminosity (L$_{Edd}$) $\simeq$ 1.3 $\times$ 10$^{38}$ (M$_{BH}$/M$_{\odot}$), representing an upper limit to the luminosity produced by a BH having mass M$_{BH}$. This limit decides the balance between the outward radiation pressure from the accreting matter and the inward gravitational pressure exerted by the BH. Here, the accreting mass rate can be parameterized by Eddington ratio ($l_{Edd}$ = L$_{Bol}$/L$_{Edd}$) which is the ratio of the bolometric luminosity (L$_{Bol}$) to the Eddington luminosity (L$_{Edd}$) for a given BH mass (M$_{BH}$). The Eddington ratio can be given as \citep{WangJ2007}

\begin{equation}
    l_{Edd} = 0.1 (L_{Bol}/1.4 \times 10^{44} erg~s^{-1}) (M_{BH}/M_{\odot}).
\end{equation}

We observationally constrain the bolometric luminosity using [OIII] $\lambda$5007 line luminosity observed in the MaNGA emission line map and the relation L$_{Bol}$/L$_{[OIII] \lambda 5007}$ $\simeq$ 3500 \citep{Heckman2004}. Since MJ0818+2257 falls in the category of Seyfert type-2 class as observationally constrained by our BPT diagnostic (see Fig.~\ref{fig-2}), we therefore assume its Eddington ratio ($l_{Edd}$) spanned over $\sim$ 1.9 $\times$ 10$^{-3}$ to $\sim$ 6.2 $\times$ 10$^{-1}$ with the median value $\sim$ 4.2 $\times$ 10$^{-2}$ \citep{Singh2011}. With this assumption and using Eq. (3) and median value of $l_{Edd}$, we reach to the BH mass hosted by MJ0818+2257 as $\sim$ 6.5 $\times$ 10$^{6}$ M$_{\odot}$. Assuming an upper limit of the galaxy stellar mass to be  M(*)$<3\times10^9$M$_{\odot}$ (see Appendix A), the SMBH mass to stellar mass ratio is M(BH)/M(*)$>$0.022. This is slightly higher than the local value of $\sim$0.001.

\bibliography{ms_AGN_LSBdwarfG}

\begin{thebibliography}{}
\expandafter\ifx\csname natexlab\endcsname\relax\def\natexlab#1{#1}\fi
\providecommand{\url}[1]{\href{#1}{#1}}
\providecommand{\dodoi}[1]{doi:~\href{http://doi.org/#1}{\nolinkurl{#1}}}
\providecommand{\doeprint}[1]{\href{http://ascl.net/#1}{\nolinkurl{http://ascl.net/#1}}}
\providecommand{\doarXiv}[1]{\href{https://arxiv.org/abs/#1}{\nolinkurl{https://arxiv.org/abs/#1}}}

\bibitem[{{Adami} {et~al.}(2006){Adami}, {Scheidegger}, {Ulmer}, {Durret},
  {Mazure}, {West}, {Conselice}, {Gregg}, {Kasun}, {Pell{\'o}}, \&
  {Picat}}]{Adami2006}
{Adami}, C., {Scheidegger}, R., {Ulmer}, M., {et~al.} 2006, \aap, 459, 679,
  \dodoi{10.1051/0004-6361:20053758}

\bibitem[{{Angl{\'e}s-Alc{\'a}zar} {et~al.}(2017){Angl{\'e}s-Alc{\'a}zar},
  {Faucher-Gigu{\`e}re}, {Quataert}, {Hopkins}, {Feldmann}, {Torrey}, {Wetzel},
  \& {Kere{\v{s}}}}]{Angles-Alcazar2017}
{Angl{\'e}s-Alc{\'a}zar}, D., {Faucher-Gigu{\`e}re}, C.-A., {Quataert}, E.,
  {et~al.} 2017, \mnras, 472, L109, \dodoi{10.1093/mnrasl/slx161}

\bibitem[{{Atek} {et~al.}(2024){Atek}, {Labb{\'e}}, {Furtak}, {Chemerynska},
  {Fujimoto}, {Setton}, {Miller}, {Oesch}, {Bezanson}, {Price}, {Dayal},
  {Zitrin}, {Kokorev}, {Weaver}, {Brammer}, {Dokkum}, {Williams}, {Cutler},
  {Feldmann}, {Fudamoto}, {Greene}, {Leja}, {Maseda}, {Muzzin}, {Pan},
  {Papovich}, {Nelson}, {Nanayakkara}, {Stark}, {Stefanon}, {Suess}, {Wang}, \&
  {Whitaker}}]{atek.etal.2024}
{Atek}, H., {Labb{\'e}}, I., {Furtak}, L.~J., {et~al.} 2024, \nat, 626, 975,
  \dodoi{10.1038/s41586-024-07043-6}

\bibitem[{{Bait} {et~al.}(2019){Bait}, {Wadadekar}, \& {Barway}}]{Omkar2019}
{Bait}, O., {Wadadekar}, Y., \& {Barway}, S. 2019, \mnras, 485, 428,
  \dodoi{10.1093/mnras/stz433}

\bibitem[{{Baldassare} {et~al.}(2015){Baldassare}, {Reines}, {Gallo}, \&
  {Greene}}]{Baldassare2015}
{Baldassare}, V.~F., {Reines}, A.~E., {Gallo}, E., \& {Greene}, J.~E. 2015,
  \apjl, 809, L14, \dodoi{10.1088/2041-8205/809/1/L14}

\bibitem[{{Baldwin} {et~al.}(1981){Baldwin}, {Phillips}, \&
  {Terlevich}}]{Baldwin1981}
{Baldwin}, J.~A., {Phillips}, M.~M., \& {Terlevich}, R. 1981, \pasp, 93, 5,
  \dodoi{10.1086/130766}

\bibitem[{{Becker} {et~al.}(1995){Becker}, {White}, \& {Helfand}}]{Becker1995}
{Becker}, R.~H., {White}, R.~L., \& {Helfand}, D.~J. 1995, \apj, 450, 559,
  \dodoi{10.1086/176166}

\bibitem[{{Bell} \& {de Jong}(2001)}]{Bell2001}
{Bell}, E.~F., \& {de Jong}, R.~S. 2001, \apj, 550, 212, \dodoi{10.1086/319728}

\bibitem[{{Bell} {et~al.}(2003){Bell}, {McIntosh}, {Katz}, \&
  {Weinberg}}]{Bell2003}
{Bell}, E.~F., {McIntosh}, D.~H., {Katz}, N., \& {Weinberg}, M.~D. 2003, \apjs,
  149, 289, \dodoi{10.1086/378847}

\bibitem[{{Bellovary} {et~al.}(2011){Bellovary}, {Volonteri}, {Governato},
  {Shen}, {Quinn}, \& {Wadsley}}]{Bellovary2011}
{Bellovary}, J., {Volonteri}, M., {Governato}, F., {et~al.} 2011, \apj, 742,
  13, \dodoi{10.1088/0004-637X/742/1/13}

\bibitem[{{Bentz} \& {Manne-Nicholas}(2018)}]{Bentz2018}
{Bentz}, M.~C., \& {Manne-Nicholas}, E. 2018, \apj, 864, 146,
  \dodoi{10.3847/1538-4357/aad808}

\bibitem[{{Bogd{\'a}n} {et~al.}(2023){Bogd{\'a}n}, {Goulding}, {Natarajan},
  {Kov{\'a}cs}, {Tremblay}, {Chadayammuri}, {Volonteri}, {Kraft}, {Forman},
  {Jones}, {Churazov}, \& {Zhuravleva}}]{Bogdan2023}
{Bogd{\'a}n}, {\'A}., {Goulding}, A.~D., {Natarajan}, P., {et~al.} 2023, Nature
  Astronomy, \dodoi{10.1038/s41550-023-02111-9}

\bibitem[{{Bundy} {et~al.}(2015){Bundy}, {Bershady}, {Law}, {Yan}, {Drory},
  {MacDonald}, {Wake}, {Cherinka}, {S{\'a}nchez-Gallego}, {Weijmans}, {Thomas},
  {Tremonti}, {Masters}, {Coccato}, {Diamond-Stanic}, {Arag{\'o}n-Salamanca},
  {Avila-Reese}, {Badenes}, {Falc{\'o}n-Barroso}, {Belfiore}, {Bizyaev},
  {Blanc}, {Bland-Hawthorn}, {Blanton}, {Brownstein}, {Byler}, {Cappellari},
  {Conroy}, {Dutton}, {Emsellem}, {Etherington}, {Frinchaboy}, {Fu}, {Gunn},
  {Harding}, {Johnston}, {Kauffmann}, {Kinemuchi}, {Klaene}, {Knapen},
  {Leauthaud}, {Li}, {Lin}, {Maiolino}, {Malanushenko}, {Malanushenko}, {Mao},
  {Maraston}, {McDermid}, {Merrifield}, {Nichol}, {Oravetz}, {Pan}, {Parejko},
  {Sanchez}, {Schlegel}, {Simmons}, {Steele}, {Steinmetz}, {Thanjavur},
  {Thompson}, {Tinker}, {van den Bosch}, {Westfall}, {Wilkinson}, {Wright},
  {Xiao}, \& {Zhang}}]{Bundy2015}
{Bundy}, K., {Bershady}, M.~A., {Law}, D.~R., {et~al.} 2015, \apj, 798, 7,
  \dodoi{10.1088/0004-637X/798/1/7}

\bibitem[{{Cammelli} {et~al.}(2024){Cammelli}, {Monaco}, {Tan}, {Singh},
  {Fontanot}, {De Lucia}, {Hirschmann}, \& {Xie}}]{cammelli.etal.2024}
{Cammelli}, V., {Monaco}, P., {Tan}, J.~C., {et~al.} 2024, arXiv e-prints,
  arXiv:2407.09949, \dodoi{10.48550/arXiv.2407.09949}

\bibitem[{{Cappellari} \& {Emsellem}(2004)}]{Cappellari2004}
{Cappellari}, M., \& {Emsellem}, E. 2004, \pasp, 116, 138,
  \dodoi{10.1086/381875}

\bibitem[{{Cid Fernandes} {et~al.}(2010){Cid Fernandes}, {Stasi{\'n}ska},
  {Schlickmann}, {Mateus}, {Vale Asari}, {Schoenell}, \&
  {Sodr{\'e}}}]{Fernandes2010}
{Cid Fernandes}, R., {Stasi{\'n}ska}, G., {Schlickmann}, M.~S., {et~al.} 2010,
  \mnras, 403, 1036, \dodoi{10.1111/j.1365-2966.2009.16185.x}

\bibitem[{{Conroy}(2013)}]{Conroy2013}
{Conroy}, C. 2013, \araa, 51, 393, \dodoi{10.1146/annurev-astro-082812-141017}

\bibitem[{{Courteau} {et~al.}(2014){Courteau}, {Cappellari}, {de Jong},
  {Dutton}, {Emsellem}, {Hoekstra}, {Koopmans}, {Mamon}, {Maraston}, {Treu}, \&
  {Widrow}}]{Courteau2014}
{Courteau}, S., {Cappellari}, M., {de Jong}, R.~S., {et~al.} 2014, Reviews of
  Modern Physics, 86, 47, \dodoi{10.1103/RevModPhys.86.47}

\bibitem[{{Dayal}(2024)}]{dayal.2024}
{Dayal}, P. 2024, arXiv e-prints, arXiv:2407.07162,
  \dodoi{10.48550/arXiv.2407.07162}

\bibitem[{{Drory} {et~al.}(2015){Drory}, {MacDonald}, {Bershady}, {Bundy},
  {Gunn}, {Law}, {Smith}, {Stoll}, {Tremonti}, {Wake}, {Yan}, {Weijmans},
  {Byler}, {Cherinka}, {Cope}, {Eigenbrot}, {Harding}, {Holder}, {Huehnerhoff},
  {Jaehnig}, {Jansen}, {Klaene}, {Paat}, {Percival}, \&
  {Sayres}}]{drory.etal.2015}
{Drory}, N., {MacDonald}, N., {Bershady}, M.~A., {et~al.} 2015, \aj, 149, 77,
  \dodoi{10.1088/0004-6256/149/2/77}

\bibitem[{{Ferrarese} \& {Merritt}(2000)}]{Ferrarese2000}
{Ferrarese}, L., \& {Merritt}, D. 2000, \apjl, 539, L9, \dodoi{10.1086/312838}

\bibitem[{{Filippenko} \& {Ho}(2003)}]{Filippenko2003}
{Filippenko}, A.~V., \& {Ho}, L.~C. 2003, \apjl, 588, L13,
  \dodoi{10.1086/375361}

\bibitem[{{Gallazzi} \& {Bell}(2009)}]{Gallazzi2009}
{Gallazzi}, A., \& {Bell}, E.~F. 2009, \apjs, 185, 253,
  \dodoi{10.1088/0067-0049/185/2/253}

\bibitem[{{Gebhardt} {et~al.}(2000){Gebhardt}, {Bender}, {Bower}, {Dressler},
  {Faber}, {Filippenko}, {Green}, {Grillmair}, {Ho}, {Kormendy}, {Lauer},
  {Magorrian}, {Pinkney}, {Richstone}, \& {Tremaine}}]{Gebhardt2000}
{Gebhardt}, K., {Bender}, R., {Bower}, G., {et~al.} 2000, \apjl, 539, L13,
  \dodoi{10.1086/312840}

\bibitem[{{Greene}(2012)}]{Greene2012}
{Greene}, J.~E. 2012, Nature Communications, 3, 1304,
  \dodoi{10.1038/ncomms2314}

\bibitem[{{Greene} \& {Ho}(2007)}]{Greene2007}
{Greene}, J.~E., \& {Ho}, L.~C. 2007, \apj, 670, 92, \dodoi{10.1086/522082}

\bibitem[{{Gunawardhana} {et~al.}(2013){Gunawardhana}, {Hopkins},
  {Bland-Hawthorn}, {Brough}, {Sharp}, {Loveday}, {Taylor}, {Jones},
  {Lara-L{\'o}pez}, {Bauer}, {Colless}, {Owers}, {Baldry},
  {L{\'o}pez-S{\'a}nchez}, {Foster}, {Bamford}, {Brown}, {Driver},
  {Drinkwater}, {Liske}, {Meyer}, {Norberg}, {Robotham}, {Ching}, {Cluver},
  {Croom}, {Kelvin}, {Prescott}, {Steele}, {Thomas}, \&
  {Wang}}]{Gunawardhana2013}
{Gunawardhana}, M.~L.~P., {Hopkins}, A.~M., {Bland-Hawthorn}, J., {et~al.}
  2013, \mnras, 433, 2764, \dodoi{10.1093/mnras/stt890}

\bibitem[{{Habouzit} {et~al.}(2017){Habouzit}, {Volonteri}, \&
  {Dubois}}]{Habouzit2017}
{Habouzit}, M., {Volonteri}, M., \& {Dubois}, Y. 2017, \mnras, 468, 3935,
  \dodoi{10.1093/mnras/stx666}

\bibitem[{{Heckman} {et~al.}(2004){Heckman}, {Kauffmann}, {Brinchmann},
  {Charlot}, {Tremonti}, \& {White}}]{Heckman2004}
{Heckman}, T.~M., {Kauffmann}, G., {Brinchmann}, J., {et~al.} 2004, \apj, 613,
  109, \dodoi{10.1086/422872}

\bibitem[{{Honey} {et~al.}(2018){Honey}, {van Driel}, {Das}, \&
  {Martin}}]{honey.etal.2018}
{Honey}, M., {van Driel}, W., {Das}, M., \& {Martin}, J.~M. 2018, \mnras, 476,
  4488, \dodoi{10.1093/mnras/sty530}

\bibitem[{{J{\'o}zsa} {et~al.}(2009){J{\'o}zsa}, {Garrett}, {Oosterloo},
  {Rampadarath}, {Paragi}, {van Arkel}, {Lintott}, {Keel}, {Schawinski}, \&
  {Edmondson}}]{Jozsa2009}
{J{\'o}zsa}, G.~I.~G., {Garrett}, M.~A., {Oosterloo}, T.~A., {et~al.} 2009,
  \aap, 500, L33, \dodoi{10.1051/0004-6361/200912402}

\bibitem[{{Kormendy} \& {Ho}(2013)}]{Kormendy2013}
{Kormendy}, J., \& {Ho}, L.~C. 2013, \araa, 51, 511,
  \dodoi{10.1146/annurev-astro-082708-101811}

\bibitem[{{Lacerda} {et~al.}(2018){Lacerda}, {Cid Fernandes}, {Couto},
  {Stasi{\'n}ska}, {Garc{\'\i}a-Benito}, {Vale Asari}, {P{\'e}rez},
  {Gonz{\'a}lez Delgado}, {S{\'a}nchez}, \& {de Amorim}}]{Lacerda2018}
{Lacerda}, E.~A.~D., {Cid Fernandes}, R., {Couto}, G.~S., {et~al.} 2018,
  \mnras, 474, 3727, \dodoi{10.1093/mnras/stx3022}

\bibitem[{{Lacy} {et~al.}(2020){Lacy}, {Baum}, {Chandler}, {Chatterjee},
  {Clarke}, {Deustua}, {English}, {Farnes}, {Gaensler}, {Gugliucci},
  {Hallinan}, {Kent}, {Kimball}, {Law}, {Lazio}, {Marvil}, {Mao}, {Medlin},
  {Mooley}, {Murphy}, {Myers}, {Osten}, {Richards}, {Rosolowsky}, {Rudnick},
  {Schinzel}, {Sivakoff}, {Sjouwerman}, {Taylor}, {White}, {Wrobel},
  {Andernach}, {Beasley}, {Berger}, {Bhatnager}, {Birkinshaw}, {Bower},
  {Brandt}, {Brown}, {Burke-Spolaor}, {Butler}, {Comerford}, {Demorest}, {Fu},
  {Giacintucci}, {Golap}, {G{\"u}th}, {Hales}, {Hiriart}, {Hodge}, {Horesh},
  {Ivezi{\'c}}, {Jarvis}, {Kamble}, {Kassim}, {Liu}, {Loinard}, {Lyons},
  {Masters}, {Mezcua}, {Moellenbrock}, {Mroczkowski}, {Nyland}, {O'Dea},
  {O'Sullivan}, {Peters}, {Radford}, {Rao}, {Robnett}, {Salcido}, {Shen},
  {Sobotka}, {Witz}, {Vaccari}, {van Weeren}, {Vargas}, {Williams}, \&
  {Yoon}}]{Lacy2020}
{Lacy}, M., {Baum}, S.~A., {Chandler}, C.~J., {et~al.} 2020, \pasp, 132,
  035001, \dodoi{10.1088/1538-3873/ab63eb}

\bibitem[{{Law} {et~al.}(2016){Law}, {Cherinka}, {Yan}, {Andrews}, {Bershady},
  {Bizyaev}, {Blanc}, {Blanton}, {Bolton}, {Brownstein}, {Bundy}, {Chen},
  {Drory}, {D'Souza}, {Fu}, {Jones}, {Kauffmann}, {MacDonald}, {Masters},
  {Newman}, {Parejko}, {S{\'a}nchez-Gallego}, {S{\'a}nchez}, {Schlegel},
  {Thomas}, {Wake}, {Weijmans}, {Westfall}, \& {Zhang}}]{Law2016}
{Law}, D.~R., {Cherinka}, B., {Yan}, R., {et~al.} 2016, \aj, 152, 83,
  \dodoi{10.3847/0004-6256/152/4/83}

\bibitem[{{Marconi} \& {Hunt}(2003)}]{Marconi2003}
{Marconi}, A., \& {Hunt}, L.~K. 2003, \apjl, 589, L21, \dodoi{10.1086/375804}

\bibitem[{{McConnell} \& {Ma}(2013)}]{McConnell2013}
{McConnell}, N.~J., \& {Ma}, C.-P. 2013, \apj, 764, 184,
  \dodoi{10.1088/0004-637X/764/2/184}

\bibitem[{{McGaugh} \& {Schombert}(2014)}]{McGaugh2014}
{McGaugh}, S.~S., \& {Schombert}, J.~M. 2014, \aj, 148, 77,
  \dodoi{10.1088/0004-6256/148/5/77}

\bibitem[{{Mezcua} \& {Dom{\'\i}nguez S{\'a}nchez}(2020)}]{Mezcua2020}
{Mezcua}, M., \& {Dom{\'\i}nguez S{\'a}nchez}, H. 2020, \apjl, 898, L30,
  \dodoi{10.3847/2041-8213/aba199}

\bibitem[{{O'Donnell}(1994)}]{O'Donnell1994}
{O'Donnell}, J.~E. 1994, \apj, 422, 158, \dodoi{10.1086/173713}

\bibitem[{{Oke}(1974)}]{Oke1974}
{Oke}, J.~B. 1974, \apjs, 27, 21, \dodoi{10.1086/190287}

\bibitem[{{Pacucci} {et~al.}(2023){Pacucci}, {Nguyen}, {Carniani}, {Maiolino},
  \& {Fan}}]{pacucci.etal.2023}
{Pacucci}, F., {Nguyen}, B., {Carniani}, S., {Maiolino}, R., \& {Fan}, X. 2023,
  \apjl, 957, L3, \dodoi{10.3847/2041-8213/ad0158}

\bibitem[{{Paswan} {et~al.}(2018){Paswan}, {Omar}, \& {Jaiswal}}]{Paswan2018}
{Paswan}, A., {Omar}, A., \& {Jaiswal}, S. 2018, \mnras, 473, 4566,
  \dodoi{10.1093/mnras/stx2614}

\bibitem[{{Paswan} {et~al.}(2022){Paswan}, {Saha}, {Borgohain}, {Leitherer}, \&
  {Dhiwar}}]{Paswan2022}
{Paswan}, A., {Saha}, K., {Borgohain}, A., {Leitherer}, C., \& {Dhiwar}, S.
  2022, \apj, 929, 50, \dodoi{10.3847/1538-4357/ac5c4b}

\bibitem[{{Ramya} {et~al.}(2011){Ramya}, {Prabhu}, \& {Das}}]{Ramya2011}
{Ramya}, S., {Prabhu}, T.~P., \& {Das}, M. 2011, \mnras, 418, 789,
  \dodoi{10.1111/j.1365-2966.2011.19530.x}

\bibitem[{{Ramya} {et~al.}(2009){Ramya}, {Sahu}, \& {Prabhu}}]{Ramya2009}
{Ramya}, S., {Sahu}, D.~K., \& {Prabhu}, T.~P. 2009, \mnras, 396, 97,
  \dodoi{10.1111/j.1365-2966.2009.14731.x}

\bibitem[{{Reines} {et~al.}(2020){Reines}, {Condon}, {Darling}, \&
  {Greene}}]{Reines2020}
{Reines}, A.~E., {Condon}, J.~J., {Darling}, J., \& {Greene}, J.~E. 2020, \apj,
  888, 36, \dodoi{10.3847/1538-4357/ab4999}

\bibitem[{{Reines} \& {Deller}(2012)}]{Reines2012}
{Reines}, A.~E., \& {Deller}, A.~T. 2012, \apjl, 750, L24,
  \dodoi{10.1088/2041-8205/750/1/L24}

\bibitem[{{Reines} {et~al.}(2013){Reines}, {Greene}, \& {Geha}}]{Reines2013}
{Reines}, A.~E., {Greene}, J.~E., \& {Geha}, M. 2013, \apj, 775, 116,
  \dodoi{10.1088/0004-637X/775/2/116}

\bibitem[{{Reines} {et~al.}(2016){Reines}, {Reynolds}, {Miller}, {Sivakoff},
  {Greene}, {Hickox}, \& {Johnson}}]{Reines2016}
{Reines}, A.~E., {Reynolds}, M.~T., {Miller}, J.~M., {et~al.} 2016, \apjl, 830,
  L35, \dodoi{10.3847/2041-8205/830/2/L35}

\bibitem[{{Reines} {et~al.}(2011){Reines}, {Sivakoff}, {Johnson}, \&
  {Brogan}}]{Reines2011Nature}
{Reines}, A.~E., {Sivakoff}, G.~R., {Johnson}, K.~E., \& {Brogan}, C.~L. 2011,
  \nat, 470, 66, \dodoi{10.1038/nature09724}

\bibitem[{{Roediger} \& {Courteau}(2015)}]{Roediger2015}
{Roediger}, J.~C., \& {Courteau}, S. 2015, \mnras, 452, 3209,
  \dodoi{10.1093/mnras/stv1499}

\bibitem[{{Schawinski} {et~al.}(2010){Schawinski}, {Evans}, {Virani}, {Urry},
  {Keel}, {Natarajan}, {Lintott}, {Manning}, {Coppi}, {Kaviraj}, {Bamford},
  {J{\'o}zsa}, {Garrett}, {van Arkel}, {Gay}, \& {Fortson}}]{Schawinski2010}
{Schawinski}, K., {Evans}, D.~A., {Virani}, S., {et~al.} 2010, \apjl, 724, L30,
  \dodoi{10.1088/2041-8205/724/1/L30}

\bibitem[{{Schutte} {et~al.}(2019){Schutte}, {Reines}, \&
  {Greene}}]{Schutte2019}
{Schutte}, Z., {Reines}, A.~E., \& {Greene}, J.~E. 2019, \apj, 887, 245,
  \dodoi{10.3847/1538-4357/ab35dd}

\bibitem[{{Seth} {et~al.}(2014){Seth}, {van den Bosch}, {Mieske}, {Baumgardt},
  {Brok}, {Strader}, {Neumayer}, {Chilingarian}, {Hilker}, {McDermid},
  {Spitler}, {Brodie}, {Frank}, \& {Walsh}}]{Anil2014}
{Seth}, A.~C., {van den Bosch}, R., {Mieske}, S., {et~al.} 2014, \nat, 513,
  398, \dodoi{10.1038/nature13762}

\bibitem[{{Singh} {et~al.}(2011){Singh}, {Shastri}, \& {Risaliti}}]{Singh2011}
{Singh}, V., {Shastri}, P., \& {Risaliti}, G. 2011, \aap, 533, A128,
  \dodoi{10.1051/0004-6361/201117422}

\bibitem[{{Subramanian} {et~al.}(2016){Subramanian}, {Ramya}, {Das}, {George},
  {Sivarani}, \& {Prabhu}}]{subramanian.etal.2016}
{Subramanian}, S., {Ramya}, S., {Das}, M., {et~al.} 2016, \mnras, 455, 3148,
  \dodoi{10.1093/mnras/stv2500}

\bibitem[{{Tremonti} {et~al.}(2004){Tremonti}, {Heckman}, {Kauffmann},
  {Brinchmann}, {Charlot}, {White}, {Seibert}, {Peng}, {Schlegel}, {Uomoto},
  {Fukugita}, \& {Brinkmann}}]{Tremonti2004}
{Tremonti}, C.~A., {Heckman}, T.~M., {Kauffmann}, G., {et~al.} 2004, \apj, 613,
  898, \dodoi{10.1086/423264}

\bibitem[{{Volonteri}(2010)}]{Volonteri2010}
{Volonteri}, M. 2010, \aapr, 18, 279, \dodoi{10.1007/s00159-010-0029-x}

\bibitem[{{Walcher} {et~al.}(2011){Walcher}, {Groves}, {Budav{\'a}ri}, \&
  {Dale}}]{Walcher2011}
{Walcher}, J., {Groves}, B., {Budav{\'a}ri}, T., \& {Dale}, D. 2011, \apss,
  331, 1, \dodoi{10.1007/s10509-010-0458-z}

\bibitem[{{Wang} \& {Zhang}(2007)}]{WangJ2007}
{Wang}, J.-M., \& {Zhang}, E.-P. 2007, \apj, 660, 1072, \dodoi{10.1086/513685}

\bibitem[{{Westfall} {et~al.}(2019){Westfall}, {Cappellari}, {Bershady},
  {Bundy}, {Belfiore}, {Ji}, {Law}, {Schaefer}, {Shetty}, {Tremonti}, {Yan},
  {Andrews}, {Brownstein}, {Cherinka}, {Coccato}, {Drory}, {Maraston},
  {Parikh}, {S{\'a}nchez-Gallego}, {Thomas}, {Weijmans}, {Barrera-Ballesteros},
  {Du}, {Goddard}, {Li}, {Masters}, {Ibarra Medel}, {S{\'a}nchez}, {Yang},
  {Zheng}, \& {Zhou}}]{Westfall2019}
{Westfall}, K.~B., {Cappellari}, M., {Bershady}, M.~A., {et~al.} 2019, \aj,
  158, 231, \dodoi{10.3847/1538-3881/ab44a2}

\bibitem[{{Yadav} {et~al.}(2022){Yadav}, {Das}, {Barway}, \&
  {Combes}}]{yadav.etal.2022}
{Yadav}, J., {Das}, M., {Barway}, S., \& {Combes}, F. 2022, \aap, 657, L10,
  \dodoi{10.1051/0004-6361/202142477}

\bibitem[{{Zhang} {et~al.}(2017){Zhang}, {Yan}, {Bundy}, {Bershady}, {Haffner},
  {Walterbos}, {Maiolino}, {Tremonti}, {Thomas}, {Drory}, {Jones}, {Belfiore},
  {S{\'a}nchez}, {Diamond-Stanic}, {Bizyaev}, {Nitschelm}, {Andrews},
  {Brinkmann}, {Brownstein}, {Cheung}, {Li}, {Law}, {Roman Lopes}, {Oravetz},
  {Pan}, {Storchi Bergmann}, \& {Simmons}}]{Zhang2017}
{Zhang}, K., {Yan}, R., {Bundy}, K., {et~al.} 2017, \mnras, 466, 3217,
  \dodoi{10.1093/mnras/stw3308}

\end{thebibliography}

\end{document}